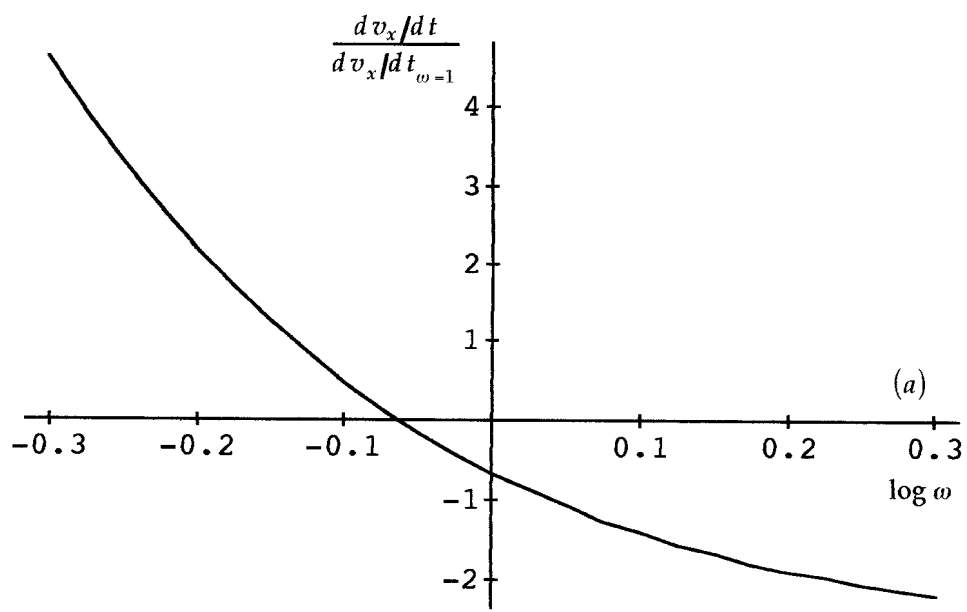

(a)

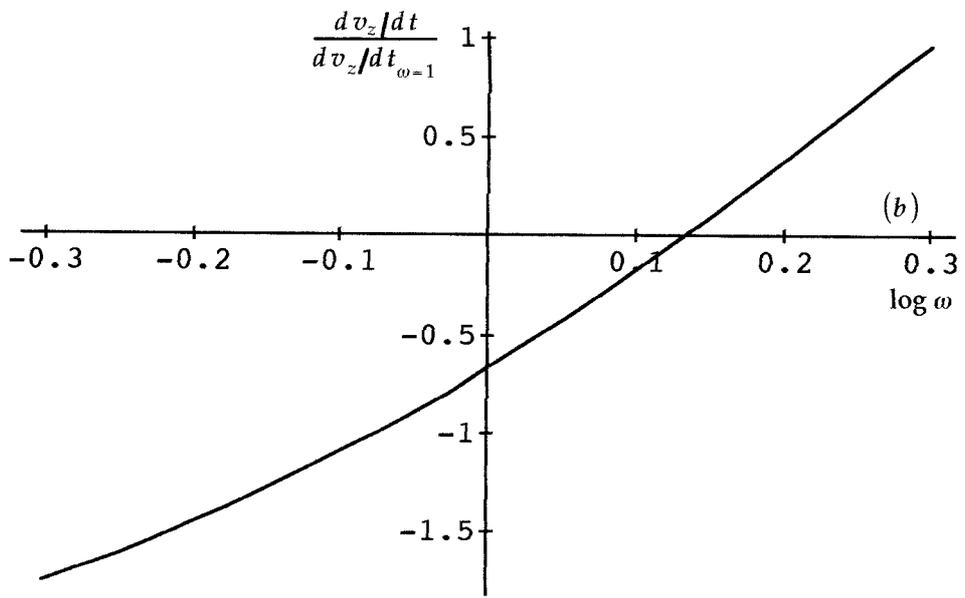

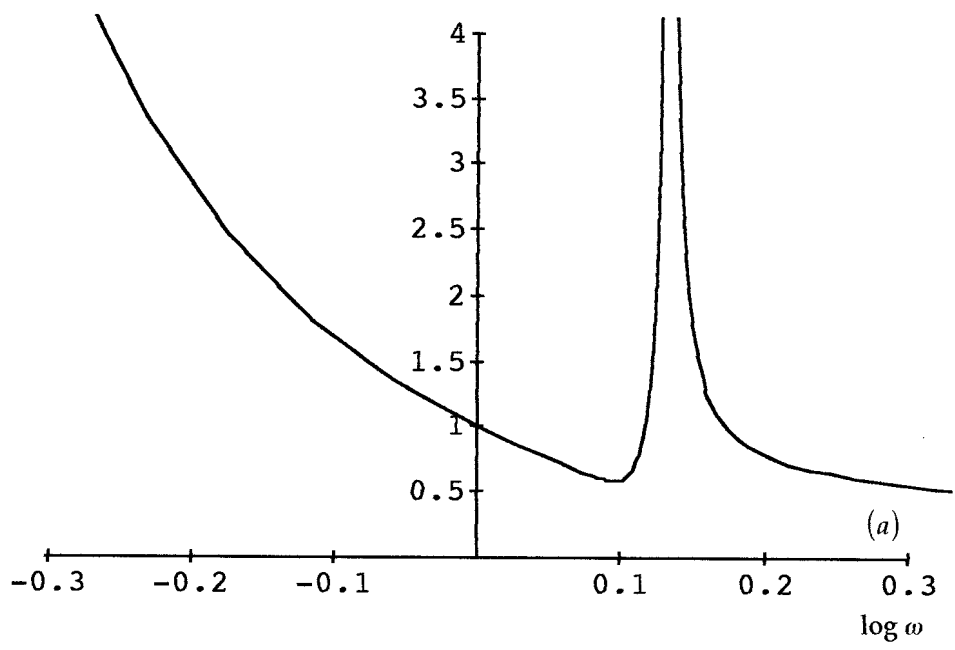

(a)

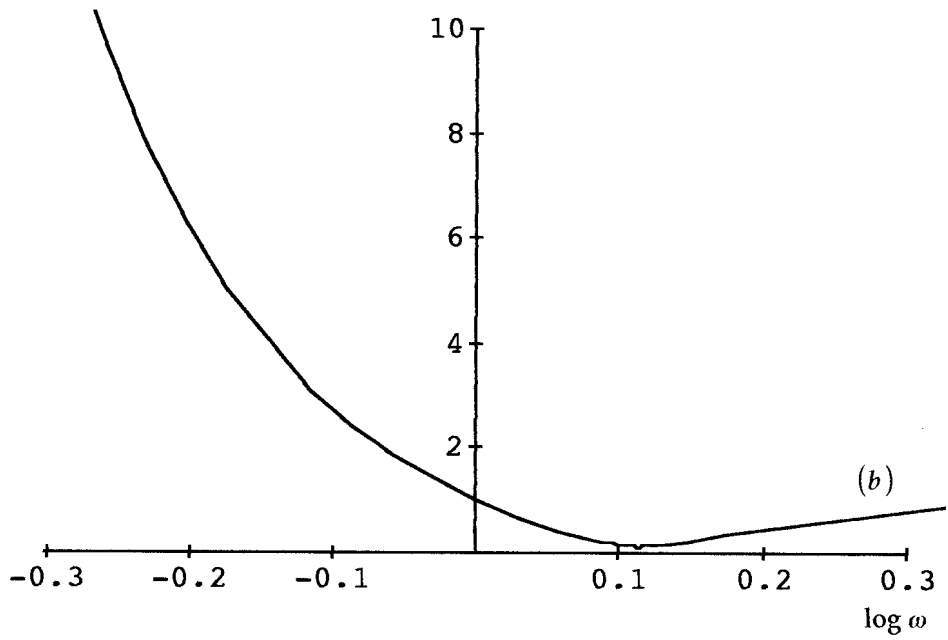
(b)

**Figure 2**: The noise level for an imperfect particle distribution for an ellipsoid kernel relative a spherical one. In Fig. 2a the corrected error, Eq. 28, is plotted, and in Fig. 2b the uncorrected, Eq. 29.

# The Use of Non-Spherical Kernels in Smooth Particle Hydrodynamics

M. Selhammar
Uppsala Astronomical Observatory
Box 515
S-751 21 Uppsala
magnus@astro.uu.se

## Abstract

A general expression for the momentum equation in the Smooth Particle Hydrodynamics approximation is derived for an arbitrary shape of the kernel. The expression is specialized to an ellipsoidal kernel, and compared with its spherical counterpart for various degrees of ellipsoidity. For such an ellipsoidal kernel the contribution to the adiabatic acceleration from different parts of the kernel is studied, and how the noise level in an interaction varies with the shape of the kernel. The two body interaction involving kernels with different shapes is also briefly discussed, as well as the conservation problems in the sum over the two body interactions.

## 1. Introduction

In Smooth Particle Hydrodynamics, (SPH), the hydrodynamical quantities are calculated for a particle and a number of other particles in its neighbourhood, called the neighbours. This neighbourhood is defined by a function called the kernel, with an extension in space delimited by the smoothing length, $h$, which confines the neighbours to within a certain volume. All lagrangian character of the particles change its set of neighbours each iteration as they move around in space. This makes it easy to model three dimensional phenomena and avoid mesh tangling problems. The particle formulation also makes it possible to describe large density differences in the same model without wasting computer resources on empty volumes. The density gradients may be steep in some cases, with a length scale of the order of a smoothing length. This may give problems in for example modelling a disk, where higher resolution is desired in the vertical rather than the horizontal direction.

In discussions about SPH, it is sometimes suggested the use of non-spherical kernels could offer an improvement of the method. If the kernel is made non-spherical and is allowed to attain a shape that is adaptive to its surrounding, according to some method, it would be possible to give the method an adaptive resolution for each particle individually. In the case of a disk it is desirable with higher resolution in the vertical direction, and therefore it could be suitable using an ellipsoidal kernel. Martel et. al. (1994) presented an implementation of non-spherical kernels, called Adaptive SPH, (ASPH), which they in Shapiro et. al. (1996) used in applications to of cosmological collapse problems. The smoothing length is an anisotropic tensor, **H**, which describes the shape of the kernel. The artificial viscosity is often a problem since it, for example, prevents desirable compression in the model. **H** is therefore also used to trace shocks in the model and reduce the use of artificial viscosity. Shapiro et. al. argue that their method gives better results than the standard formulation of SPH, particularly when the models involve large anisotropic density changes and strong shocks. Further ASPH can give structures with few particles, and avoid some of the errouneous artificial viscous heating, that is often a problem in collapse simulations.

Fulbright et. al. (1995) present a similar description of a SPH implementation with non-sphererical kernels. They show that spherical kernels cause problems when gas is compressed or when a star is torn apart by a black hole, and argue that their spheroidal kernel gives better results.

Neither Martel et. al. (1994) nor Fulbright et. al. (1995) study the continuum limit of the method for a non-spherical kernel, nor do they discuss possible problems when the neighbours are not well distributed. Further Fulbright et. al. (1995) presume that the neighbours are distributed highly anisotropically in a compression, without discussing its implications. How the neighbours are distributed is essential for the accuracy of the method, including the case with a spherical kernel, and will be discussed later. Here it is presumed that a linear expansion is a good approximation of the density, according to the assumptions in the derivations of the SPH approximations from for example Benz (1990).

The SPH methodology is briefly described below. More comprehensive introductions to the method are found in for example Hernquist & Katz (1989), Monaghan (1992) or Steinmetz & Müller (1993). From the momentum equation and standard assumptions in the SPH method, an expression for the adiabatic acceleation is derived for the continuous case. This and its discrete particle version is used to study variations in the ellipsoidicity of the kernel.

## 2. The SPH Method

In SPH the fluid is modelled by particles that carry necessary physical quantities. Other quantities are calculated as a sum over the particles in a volume, described by the smoothing length, $h$. To simulate interactions between particles, the particle mass is distributed over this volume, called the kernel. Within $2h$ there are a number of other neighbours, and $h$ is often varied so that the numbers of neighbours is kept constant. Here this is less important, because mainly a continuous fluid will be considered. The smoothing length should therefore rather be seen as a measure of a typical length scale in the model. At $x^k$ a quantity can be approximated by an integration over space:

$$\langle f_{x^k} \rangle = \oint d^3 x' f_{x'^k} w_{x'^k - x^k}, \tag{1}$$

where $w$ is the so called kernel. The momentum equation:

$$\frac{dv^k}{dt} = -\frac{1}{\rho}\frac{\partial p}{\partial x^k}. \tag{2}$$

which from Eq. (1) around $x^k$ gives the SPH approximation of the momentum equation:

$$\frac{dv^k}{dt} \approx \left\langle \frac{\partial v^k}{\partial t} \right\rangle = -\left\langle \frac{\partial}{\partial x^k}\frac{p}{\rho} + \frac{p}{\rho^2}\frac{\partial \rho}{\partial x^k} \right\rangle = -\frac{\partial}{\partial x^k}\left\langle \frac{p}{\rho} \right\rangle - \frac{p}{\rho^2}\frac{\partial \langle \rho \rangle}{\partial x^k} =$$

$$-\frac{\partial}{\partial x^k}\oint d^3 x' \frac{p_{x'^k}}{\rho_{x'^k}} w_{x'^k - x^k} - \frac{p_{x^k}}{\rho_{x^k}^2}\frac{\partial}{\partial x^k}\oint d^3 x' \rho_{x'^k} w_{x'^k - x^k} =$$

$$\oint d^3 x' \left( \frac{p_{x'^k}}{\rho_{x'^k}} + \frac{p_{x^k}}{\rho_{x^k}^2} \rho_{x'^k} \right) \frac{\partial w_{x'^k - x^k}}{\partial x'^k}, \tag{3}$$

from for example Benz (1990). In Eq. 3 and hereafter the artificial viscosity is neglected, and it is also assumed that $dv^k/dt = \langle dv^k/dt \rangle$.

To calculate the integrations it is suitable to use spherical symmetry,

$$\oint d^3 x = \int_0^\infty dr \int_0^\pi d\theta \int_0^{2\pi} d\phi \, r^2 \sin\theta, \tag{4}$$

that is an integration over all space. An equation of state, $p = (\gamma - 1)u\rho$, is necessary as well, that relates the pressure, internal energy and density. This is to my knowledge the only equation of state used in SPH. This expression for the pressure is substituted into the momentum equation, Eq. 2:

$$\frac{dv^k}{dt} = -(\gamma-1)\left(\frac{\partial u}{\partial x^k} + \frac{u}{\rho}\frac{\partial \rho}{\partial x^k}\right). \tag{5}$$

In principle hte neighbours may have any distribution, but in the SPH approximation it is assumed that a linear expansion is a sufficient approximation for any involved quantity. In the SPH approximation of the momentum equation it is therefore assumed that in an arbitrary point, $x^k$,

$$\begin{cases} \rho = \rho|_0 + \left.\frac{\partial \rho}{\partial x^l}\right|_0 x^l \\ u = u|_0 + \left.\frac{\partial u}{\partial x^l}\right|_0 x^l \end{cases}. \tag{6}$$

The calculation is position independent, and therefore the integration in Eq. (3) is performed around the origin, and therefore $p_{x^k} \to p_0$ and $\rho_{x^k} \to \rho_0$, and the prime dropped on $x^k$, and the prime dropped on the remaining term. Substitution of $p = (\gamma-1)u\rho$ and a continued derivation of the adabatic term in Eq. 3 gives

$$\left.\frac{dv^k}{dt}\right|_{Ad} = (\gamma-1)\oint d^3x \left(u_{x^k} + \frac{u|_0}{\rho|_0}\rho_{x^k}\right)\frac{\partial w}{\partial x^k} =$$

$$(\gamma-1)\oint d^3x \left[u_0 + \left.\frac{\partial u}{\partial x^l}\right|_0 x^l + u_0 \frac{\rho_0 + (\partial \rho/\partial x^l)|_0 x^l}{\rho|_0}\right]\frac{\partial w}{\partial x^k} =$$

$$(\gamma-1)\left(\left.\frac{\partial u}{\partial x^l}\right|_0 + \frac{u_0}{\rho_0}\left.\frac{\partial \rho}{\partial x^l}\right|_0\right)\oint d^3x\, x^l \frac{\partial w}{\partial x^k}. \tag{7}$$

The odd terms are dropped during the calculation under the assumption that the derivative of the kernel is odd.

In all calculations the Monaghan and Lattanzio (1985) kernel is used,

$$w = \frac{1}{\pi h^3}\begin{cases} 1 - \frac{3}{2}\left(\frac{r}{h}\right)^2 + \frac{3}{4}\left(\frac{r}{h}\right)^3 & 0 \le r \le h \\ \frac{1}{4}\left(2 - \frac{r}{h}\right)^3 & h \le r \le 2h, \\ 0 & r > 2h \end{cases} \tag{8}$$

where in spherical symmetry the smoothing length is dependent on the angles $\theta$ and $\phi$. Further, the derivative of the kernel is used:

$$\frac{\partial w}{\partial x^k} = \frac{\partial w}{\partial (r/h)}\frac{\partial (r/h)}{\partial x^k} = \left(\frac{1}{h}\frac{\partial r}{\partial x^k} - \frac{r}{h^2}\frac{\partial h}{\partial x^k}\right)\frac{\partial w}{\partial (r/h)} = \left(\frac{\hat{x}^k}{h} - \frac{1}{h^2}\frac{\partial h}{\partial \hat{x}^k}\right)\frac{\partial w}{\partial (r/h)}. \tag{9}$$

The derivative of the Monaghan & Lattanzio kernel is odd, that is

$$\frac{\partial w(x^k, h)}{\partial x^k} = -\frac{\partial w(-x^k, h)}{\partial x^k}, \tag{10}$$

if $h(x^k) = h(-x^k)$, that is if $h$ is even. In Eq. 7 the odd terms can be neglected, and after substitution of Eq. 9 into it gives

$$\frac{dv^k}{dt} = (\gamma-1)\oint d^3x\, C^l x^l \frac{\partial w}{\partial x^k} = (\gamma-1)C^l \oint d^3x \left(\frac{\hat{x}^k}{h} - \frac{1}{h^2}\frac{\partial h}{\partial \hat{x}^k}\right) x^l \frac{\partial w}{\partial(r/h)} =$$

$$(\gamma-1)C^l \int_0^\pi d\theta \int_0^{2\pi} d\phi \sin\theta \left(\frac{\hat{x}^k}{h} - \frac{1}{h^2}\frac{\partial h}{\partial \hat{x}^k}\right)\hat{x}^l \int_0^\infty dr\, r^3 \frac{\partial w}{\partial(r/h)} =$$

$$-\frac{3(\gamma-1)C^l}{4\pi}\int_0^\pi d\theta \int_0^{2\pi} d\phi \sin\theta \left(\hat{x}^k \hat{x}^l - \frac{\partial h}{\partial \hat{x}^k}\frac{\hat{x}^l}{h}\right), \qquad (11)$$

where

$$C^l = \left(\left.\frac{\partial u}{\partial x^l}\right|_0 + \left.\frac{u|_0}{\rho|_0}\frac{\partial \rho}{\partial x^l}\right|_0\right). \qquad (12)$$

If the kernel is spherical, $\partial h/\partial x^k = 0$, then Eq. (11) implies that $dv^k/dt = -(\gamma-1)C^k$, which satisfies the momentum equation, Eq. (5). Hence Eq. (11) satisfies the momentum equation if and only if $\partial h/\partial x^k = 0$, and any additional term in the integrand is errouneous. Fulbright et. al. (1995) do not consider the $\partial h/\partial \hat{x}^k$-terms, and Shapiro et. al. (1996) neglect them because, as they argue, these terms are small just as they small with spherical kernels. In standard SPH $\partial h/\partial \hat{x}^k$ is should be small, because otherwise there would be conservation problems as well as possible violations of the assumptions that a linear expansion of a quantity is a sufficient approximations. Their argument should probably be taken to mean that the differences in shapes for neighbouring particles are expected to be small. Here, however, the smoothing length describes a non-spherical kernel, that is the $\partial h/\partial \hat{x}^k$-terms are not negligable for a sufficently deformed kernel, which invalidates their formulation of the SPH-approximation.

It should be emphasised that for the SPH approximations to be valid $w$ has to be an even function, and hence the calculations will be unphysical unless $h$ is even as well. Therefore, ellipsoidal kernels only are considered in the rest of this article.

# 3. An Ellipsoidal Kernel

If the kernel is an ellipsoid, its smoothing length can be written $h = \sqrt{x^l x^l}$, as a scalar product that is, from the equation of an ellipsoid,

$$\frac{x_x^2}{h_x^2} + \frac{x_y^2}{h_y^2} + \frac{x_z^2}{h_z^2} = 1, \qquad (13)$$

where the smoothing length is a function of the spherical coordinates, $h = h(h_0, \theta, \phi)$, where $h_0$ in turn is a function of the density and the number of neighbours. The volume of the ellipsoid is by definition

$$\frac{4\pi}{3} h_x h_y h_z = \frac{4\pi h_0^3}{3}\chi_x \chi_y \chi_z, \qquad (14)$$

which gives the relation $\chi_x \chi_y \chi_z = 1$, under the assumtion that the volume is constant when $h_0$ is constant. Here the ellipsoid is assumed to be symmetric in the $xy$-plane, that is

$$\chi = 1/\sqrt{\omega}, \tag{15}$$

where $\chi = \chi_x = \chi_y$ and $\omega = \chi_z$. This gives a new equation for the smoothing length,

$$h_0^2 = \omega(x^2 + y^2) + z^2/\omega^2 \tag{16}$$

In spherical symmetric coordinates the smoothing length becomes

$$h = h_0 (\omega \sin^2 \theta + \cos^2 \theta / \omega^2)^{-1/2}, \tag{17}$$

with the derivative

$$\frac{\partial h}{\partial \theta} = -\frac{h_0 \sin \theta \cos \theta (\omega - 1/\omega^2)}{(\omega \sin^2 \theta + \cos^2 \theta/\omega^2)^{3/2}}. \tag{18}$$

The smoothing length is substituted into Eq. 11, which gives

$$\frac{dv^k}{dt} = -\frac{3(\gamma-1)C^l}{4\pi} \int_0^\pi d\theta \int_0^{2\pi} d\phi \sin\theta \left( \hat{x}^k \hat{x}^l - \frac{\partial h}{\partial \theta} \frac{\partial \theta}{\partial \hat{x}^k} \frac{\hat{x}^l}{h} \right), \tag{19}$$

The contribution to the acceleration in the $x$-direction is

$$\frac{dv_x}{dt} = -\frac{3(\gamma-1)C_x}{4\pi} \int_0^\pi d\theta \int_0^{2\pi} d\phi \sin\theta \left\{ \sin^2\theta \cos^2\phi - \left[ -\frac{h_0 \sin\theta \cos\theta(\omega - 1/\omega^2)}{(\omega\sin^2\theta + \cos^2\theta/\omega^2)^{3/2}} \right] \times \right.$$

$$\frac{1}{\cos\theta \cos\phi}$$

$$-\frac{3(\gamma-1)C_x}{4\pi} \int_0^\pi d\theta \int_0^{2\pi} d\phi \sin^3\theta \left\{ \cos^2\phi + \frac{\cos\theta(\omega - 1/\omega^2)}{(\omega\sin^2\theta + \cos^2\theta/\omega^2)^{3/2}} \times \right.$$

$$\left. \frac{1}{\cos\theta\cos\phi} \frac{\cos\phi}{(\omega\sin^2\theta + \cos^2\theta/\omega^2)^{-1/2}} \right\} =$$

$$-\frac{3(\gamma-1)C_x}{4\pi} \int_0^\pi d\theta \int_0^{2\pi} d\phi \sin^3\theta \left[ \cos^2\phi + \frac{\omega - 1/\omega^2}{\omega\sin^2\theta + \cos^2\theta/\omega^2} \right] =$$

$$-\frac{3(\gamma-1)C_x}{4\pi} \int_0^\pi d\theta \sin^3\theta \left[ \pi + 2\pi \frac{\omega - 1/\omega^2}{\omega\sin^2\theta + \cos^2\theta/\omega^2} \right] =$$

$$-\frac{3(\gamma-1)C_x}{4} \int_0^\pi d\theta \sin^3\theta \left[ 1 + \frac{2(\omega^3 - 1)}{\omega^3 \sin^2\theta + \cos^2\theta} \right], \tag{20}$$

and in the $z$-direction:

$$\frac{dv_z}{dt} = -\frac{3(\gamma-1)C_z}{4\pi} \int_0^\pi d\theta \int_0^{2\pi} d\phi \sin\theta \left\{ \cos^2\theta - \left[ -\frac{h_0 \sin\theta \cos\theta(\omega - 1/\omega^2)}{(\omega \sin^2\theta + \cos^2\theta/\omega^2)^{3/2}} \frac{1}{-\sin\theta} \right] \times \right.$$

$$\left. \frac{\cos\theta}{h_0 (\omega \sin^2\theta + \cos^2\theta/\omega^2)^{-1/2}} \right\} =$$

$$-\frac{3(\gamma-1)C_z}{4\pi} \int_0^\pi d\theta \int_0^{2\pi} d\phi \sin\theta \left\{ \cos^2\theta - \left[ -\frac{h_0 \sin\theta \cos\theta(\omega - 1/\omega^2)}{(\omega \sin^2\theta + \cos^2\theta/\omega^2)^{3/2}} \frac{1}{-\sin\theta} \right] \times \right.$$

$$\left. \frac{\cos\theta}{h_0 (\omega \sin^2\theta + \cos^2\theta/\omega^2)^{-1/2}} \right\} =$$

$$-\frac{3(\gamma-1)C_z}{2} \int_0^\pi d\theta \sin\theta \cos^2\theta \left( 1 - \frac{\omega - 1/\omega^2}{\omega \sin^2\theta + \cos^2\theta/\omega^2} \right) =$$

$$-\frac{3(\gamma-1)C_z}{2} \int_0^\pi d\theta \sin\theta \cos^2\theta \left( 1 - \frac{\omega^3 - 1}{\omega^3 \sin^2\theta + \cos^2\theta} \right). \tag{21}$$

In Fig. 1 $dv_x/dt$ and $dv_z/dt$ are plotted for $\log 0.5 \leq \log \omega \leq \log 2.0$, where $\gamma = 5/3$ and $C_x = C_z = 1$. The figure shows that the acceleration does not only strongly deviate from the momentum equation for a moderate deformation of the kernel; a numerical solution gives that the acceleration even changes sign at $\omega = 0.86$ and $\omega = 1.36$ in the $x$- and $z$-direction respectively. It can be noted that $dv_x/dt \to +\infty$ if $\omega \to 0$, and asymptotically towards -2.67 if $\omega \to \infty$, while $dv_z/dt \to -2.67$ when $\omega \to 0$ and $dv_z/dt \to +\infty$ if $\omega \to \infty$. Note that for $\omega = 0$ and $\omega = \infty$ the integrands are not piecewice continuous, and therefore not integrable.

It is possible to calculate the acceleration for a known shape of the kernel, and relate it to its spherical counterpart. Hence it is, at least in principle, possible to compensate for the error in the momentum equation as a result of the non-spherical kernels. An example of a naive scaling function is

$$S^{pq} = \frac{dv^k}{dt}\bigg|_{\omega=1} \left( \frac{dv^k}{dt}\bigg|_\omega \right)^{-1} =$$

$$C^l \int_0^\pi d\theta \int_0^{2\pi} d\phi \sin\theta \, \hat{x}^p \hat{x}^l \left[ C^m \int_0^\pi d\theta \int_0^{2\pi} d\phi \sin\theta \left( \hat{x}^k \hat{x}^m - \frac{\partial h}{\partial \theta} \frac{\partial \theta}{\partial \hat{x}^k} \frac{\hat{x}^m}{h} \right) \right]^{-1}, \tag{22}$$

which gives a corrected acceleration:

$$\frac{dv^k}{dt}\bigg|_{corr} = -\frac{3(\gamma-1)C^l S^{pq} \delta_{kpqr}}{4\pi} \int_0^\pi d\theta \int_0^{2\pi} d\phi \sin\theta \left( \hat{x}^r \hat{x}^l - \frac{\partial h}{\partial \theta} \frac{\partial \theta}{\partial \hat{x}^r} \frac{\hat{x}^l}{h} \right), \tag{23}$$

where

$$\delta_{kpqr} = \begin{cases} 1 & \text{if } k = p = q = r \\ 0 & \text{otherwise} \end{cases}. \tag{24}$$

There is, however, a problem since $S^{pq}$ will go to infinity for a particular value of $\omega$. For a perfect distribution of the neighbours, the sum over them would go to zero. There are, however, always small deviations in the particles' positions, and therefore the errors become large where $S^{pq}$ is large. Therefore more sophisticated methods to compensate for the variation should be constructed.

# 4. Anisotropic Contribution to the Acceleration

In a model the distribution of the neighbours is not perfect, which gives an error in the calculation of the acceleration. Further, the contribution to the acceleration from different parts of the kernel is not isotropic. It is therefore of interest to study how this dependence of different parts of the kernel varies for different ellipsoidity.

Considering ellipsoidal kernels, dividing the integrand in Eq. 21 with

$$-\frac{3(\gamma-1)C^k}{2h_0^3}\int_0^h dr\int_0^{2\pi}d\phi\, r^2\sin\theta = -\frac{\pi(\gamma-1)h^3 C^k}{h_0^3}\sin\theta = -\frac{\pi\omega^3(\gamma-1)C^k\sin\theta}{\left(\omega^3\sin^2\theta+\cos^2\theta\right)^{3/2}}, \quad (25)$$

gives a measure of the contribution to the acceleration in the z-direction per unit volume. This gives

$$I_z = \frac{3(\gamma-1)C_z}{2}\sin\theta\cos^2\theta\left(1-\frac{\omega^3-1}{\omega^3\sin^2\theta+\cos^2\theta}\right)\frac{\left(\omega^3\sin^2\theta+\cos^2\theta\right)^{3/2}}{\pi\omega^3(\gamma-1)C_z\sin\theta} =$$

$$\frac{3\cos^2\theta\left(\omega^3\sin^2\theta+\cos^2\theta\right)^{3/2}}{2\pi\omega^3}\left(1-\frac{\omega^3-1}{\omega^3\sin^2\theta+\cos^2\theta}\right), \quad (26)$$

and from Eq. 22 the corresponding corrected expression is:

$$I_{z,\text{corr}} = S_{zz}I_z. \quad (27)$$

The noise level due to the imperfect particle distribution relative a spherical kernel can now be calculated. In the z-direction for the corrected and uncorrected contribution per unit volume, the errors are

$$E_{z,\omega,\text{corr}} = \left[\frac{S_{zz,\omega}^2\int_0^\pi d\theta\, I_{z,\omega}^2}{S_{zz,1}^2\int_0^\pi d\theta\, I_{z,1}^2}\right]^{1/2} =$$

$$\left\{\left[\int_0^\pi d\theta\sin\theta\cos^2\theta\right]^2 \times\right.$$

$$\left[\int_0^\pi d\theta\sin\theta\left(\cos^2\theta - \frac{h_0\sin\theta\cos\theta(\omega-1/\omega^2)}{\left(\omega\sin^2\theta+\cos^2\theta/\omega^2\right)^{3/2}}\frac{1}{\sin\theta}\frac{\cos\theta}{h_0\left(\omega\sin^2\theta+\cos^2\theta/\omega^2\right)^{-1/2}}\right)\right]^{-2} \times$$

$$\left[\int_0^\pi d\theta\frac{9\cos^4\theta\left(\omega^3\sin^2\theta+\cos^2\theta\right)^3}{4\pi^2\omega^6}\left(1-\frac{\omega^3-1}{\omega^3\sin^2\theta+\cos^2\theta}\right)^2\right] \times \left[\int_0^\pi d\theta\frac{9\cos^4\theta}{4\pi^2}\right]^{-1}\right\}^{1/2} =$$

$$\frac{4\pi}{3}\sqrt{\frac{8}{3\pi}}\left[\int_0^\pi d\theta \sin\theta \cos^2\theta\left(1-\frac{\omega^3-1}{\omega^3\sin^2\theta+\cos^2\theta}\right)\right]^{-1}\times$$

$$\left[\int_0^\pi d\theta \frac{\cos^4\theta(\omega^3\sin^2\theta+\cos^2\theta)^3}{\omega^6}\left(1-\frac{\omega^3-1}{\omega^3\sin^2\theta+\cos^2\theta}\right)^2\right]^{1/2}=$$

$$\frac{2}{3}\sqrt{\frac{8}{3\pi}}\left[\int_0^\pi d\theta \sin\theta \cos^2\theta\left(1-\frac{\omega^3-1}{\omega^3\sin^2\theta+\cos^2\theta}\right)\right]^{-1}\times$$

$$\left[\int_0^\pi d\theta \frac{\cos^4\theta(\omega^3\sin^2\theta+\cos^2\theta)^3}{\omega^6}\left(1-\frac{\omega^3-1}{\omega^3\sin^2\theta+\cos^2\theta}\right)^2\right]^{1/2}, \qquad (28)$$

and

$$E_{z,\omega}=\left[\frac{\int_0^\pi d\theta\, I_{z,\omega}^2}{\int_0^\pi d\theta\, I_{z,1}^2}\right]^{1/2}=$$

$$\left\{\left[\int_0^\pi d\theta \frac{9\cos^4\theta(\omega^3\sin^2\theta+\cos^2\theta)^3}{4\pi^2\omega^6}\left(1-\frac{\omega^3-1}{\omega^3\sin^2\theta+\cos^2\theta}\right)^2\right]\left[\int_0^\pi d\theta \frac{9\cos^4\theta}{4\pi^2}\right]^{-1}\right\}^{1/2}=$$

$$\left\{\left[\int_0^\pi d\theta \frac{9\cos^4\theta(\omega^3\sin^2\theta+\cos^2\theta)^3}{4\pi^2\omega^6}\left(1-\frac{\omega^3-1}{\omega^3\sin^2\theta+\cos^2\theta}\right)^2\right]\left[\frac{9}{4\pi^2}\frac{3\pi}{8}\right]^{-1}\right\}^{1/2}=$$

$$\sqrt{\frac{8}{3\pi}}\left\{\left[\int_0^\pi d\theta \frac{\cos^4\theta(\omega^3\sin^2\theta+\cos^2\theta)^3}{\omega^6}\left(1-\frac{\omega^3-1}{\omega^3\sin^2\theta+\cos^2\theta}\right)^2\right]\right\}^{1/2}. \qquad (29)$$

In Fig. 2 $E_{z,\omega,\text{corr}}$ and $E_{z,\omega}$ are plotted for $\log 0.5 \leq \log\omega \leq \log 2.0$. In Fig. 2a $E_{z,\omega,\text{corr}}$ is indefinite at $\omega=1.36$ where $dv_z/dt=0$, so obviously some other method has to be used to compensate for the variation in the acceleration if a non-sphererical kernel is used. Furthermore it shows that the error grows relatively rapidly when $\omega$ becomes small. In Fig. 2b $E_{z,\omega}$ grows fast when $\omega$ is smaller than one, which shows that when the kernel becomes flattened, fewer particles contribute to a major part of the acceleration in the z-direction. The x- and y- directions give similar relations when the kernel is elongated. This is, however, a problem only if the particle distribution gives a more than negligable error in the corresponding spherical kernel. This depends on the local particle distribution and the model as such.

## 5. The Two Body Interaction

Now we will study the particle discretization, and two nearby particles that interact with each other. From Eqs. 3 and 18 the contribution from particle j on particle i is

$$\frac{dv^k}{dt}\bigg|_{ij}=\frac{m}{\rho_j}\left(\frac{p_i}{\rho_i^2}\rho_j+\frac{p_j}{\rho_j}\right)\frac{\partial w}{\partial x^k}\bigg|_{ij}=m(\gamma-1)\left(\frac{u_i}{\rho_i}+\frac{u_j}{\rho_j}\right)\frac{\partial w}{\partial x^k}\bigg|_{ij}=$$

$$m(\gamma-1)\left(\frac{u_i}{\rho_i}+\frac{u_j}{\rho_j}\right)\frac{\partial w}{\partial(r/h)_{ij}}\frac{\partial(r/h)}{x^k}\bigg|_{ij} =$$

$$m(\gamma-1)\left(\frac{u_i}{\rho_i}+\frac{u_j}{\rho_j}\right)\left(\frac{1}{h_i}\frac{x^k_{ij}}{r_{ij}}-\frac{r_{ij}}{h_i^2}\frac{\partial h}{\partial x^k}\bigg|_i\right)\frac{\partial w}{\partial(r/h)_{ij}}. \qquad (30)$$

If $h_i = h_j$ and $\partial h/\partial x^k\big|_i = \partial h/\partial x^k\big|_j$ in Eq. 22, it is symmetric regarding $i$ and $j$, that is $dv^k/dt\big|_{ij} = dv^k/dt\big|_{ji}$. If on the other hand $h_i \neq h_j$ or $\partial h/\partial x^k\big|_i \neq \partial h/\partial x^k\big|_j$, the momentum is not conserved in the two body interaction. (There are thermal energy conservation problems as well, but the enthalpy equation is not considered here.) In the standard formulation, (where $\partial h/\partial x^k = 0$) this is usually solved using the arithmetic mean over the smoothing lengths. This does, however, introduce a lack of precision in the physical description of the fluid, because the smoothing length is a function of the density described by the environment of the particle in question.

The problem can be exemplified by the sum over $j$ in Eq. 30, where all quantities are constant except the smoothing length, that varies as

$$h = h_0 + \frac{\partial h}{\partial x^l}\bigg|_0 x^l, \qquad (31)$$

and gives

$$\frac{dv^k}{dt} = (\gamma-1)u\oint d^3x \frac{\partial w}{\partial x^k} = (\gamma-1)u\oint d^3x \left(\frac{\hat{x}^k}{h} - \frac{1}{h^2}\frac{\partial h}{\partial \hat{x}^k}\right)\frac{\partial w}{\partial(r/h)} =$$

$$(\gamma-1)u\int_0^\pi d\theta \int_0^{2\pi} d\theta \sin\theta \left(\frac{\hat{x}^k}{h} - \frac{1}{h^2}\frac{\partial h}{\partial \hat{x}^k}\right)\int_0^\infty dr\, r^2 \frac{\partial w}{\partial(r/h)} =$$

$$-\frac{7(\gamma-1)u}{10\pi}\int_0^\pi d\theta \int_0^{2\pi} d\theta \sin\theta \left(\frac{\hat{x}^k}{h} - \frac{1}{h^2}\frac{\partial h}{\partial \hat{x}^k}\right) =$$

$$-\frac{7(\gamma-1)u}{10\pi}\int_0^\pi d\theta \int_0^{2\pi} d\theta \sin\theta \left[\left(\frac{1}{h_0} - \frac{1}{h_0^2}\frac{\partial h}{\partial \hat{x}^l}\bigg|_0 \hat{x}^l\right)\hat{x}^k - \left(\frac{1}{h_0^2} - \frac{2}{h_0^3}\frac{\partial h}{\partial \hat{x}^m}\bigg|_0 \hat{x}^m\right)\frac{\partial h}{\partial \hat{x}^n}\bigg|_0 \frac{\partial \hat{x}^n}{\partial \hat{x}^k}\right] =$$

$$\frac{7(\gamma-1)u}{10\pi h_0^2}\int_0^\pi d\theta \int_0^{2\pi} d\theta \sin\theta \left(\frac{\partial h}{\partial \hat{x}^l}\bigg|_0 \hat{x}^l \hat{x}^k + \frac{\partial h}{\partial \hat{x}^k}\bigg|_0\right) = \frac{56(\gamma-1)u}{15h_0^2}\frac{\partial h}{\partial \hat{x}^k}\bigg|_0. \qquad (32)$$

From the momentum equation, Eq. 2, the acceleration is zero if the pressure is constant. If the smoothing length varies over the kernel, this is not generally true in the SPH approximation. Hence, the momentum equation may well be violated even if non-spheric kernels are introduced.

Although the smoothing length is the variable that describes the environment in the SPH approximation, it is the neighbours and their distribution around the particles that define this neighbourhood. It is assumed that a linear expansion of the physical quantities the particles carry is a good approximation. Obviously the distribution is not perfect, which induces errors in the SPH approximation as discussed by Selhammar (1997). Although the averaging of the smoothing length gives an additional error, the neighbour's distribution in itself introduces errors in the environment, and therefore it can be questioned if the exact geometrical point where the calculation is performed matters. Variation in the smoothing length may, however, give large errors, as Hernquist (1993) noted.

If the shapes of the particles $i$ and $j$ are different, the situation is more complicated and the errors possibly larger. It is not sufficient to average over the smoothing lengths, because Eq. 30 includes its derivate, $\partial h / \partial x^k$. To conserve the momentum it is possible to average over $h$ and $\partial h / \partial x^k$, that is to do the transformation

$$\left.\frac{\partial w}{\partial x^k}\right|_{ij} = \left(\frac{1}{h_i}\frac{x_{ij}^k}{r_{ij}} - \frac{r_{ij}}{h_i^2}\left.\frac{\partial h}{\partial x^k}\right|_i\right)\frac{\partial w}{\partial (r/h)_{ij}} \rightarrow \left.\frac{\partial w}{\partial x^k}\right|_{ij,\text{mean}} =$$

$$\left[\frac{1}{h_i + h_j}\frac{x_{ij}^k}{r_{ij}} - \frac{r_{ij}}{(h_i + h_j)^2}\left(\left.\frac{\partial h}{\partial x^k}\right|_i + \left.\frac{\partial h}{\partial x^k}\right|_j\right)\right]\frac{\partial w}{\partial [r_{ij}/(h_i + h_j)]}. \quad (33)$$

As discussed in Sect. 4, the calculated acceleration must be compensated to satisfy the momentum equation. As shown there were difficulties with the suggested form, but it should be sufficient for an ideal particle distribution considered here. Averaging over the smoothing lengths and its derivative gives

$$S_{ij}^{pq} =$$

$$C^l \int_0^\pi d\theta \int_0^{2\pi} d\phi \sin\theta\, \hat{x}^p \hat{x}^l \left\{ C^m \int_0^\pi d\theta \int_0^{2\pi} d\phi \sin\theta \left[ \hat{x}^q \hat{x}^m - \left(\left.\frac{\partial h}{\partial \hat{x}^q}\right|_i + \left.\frac{\partial h}{\partial \hat{x}^q}\right|_j\right)\frac{\hat{x}^m}{(h_i + h_j)}\right]\right\}^{-1}. \quad (34)$$

The acceleration for particle $i$ as a sum over the neighbours from Eqs. 31-34 becomes

$$\left.\frac{dv^k}{dt}\right|_i = \frac{m(\gamma-1)u\delta_{kpqr}}{\rho}\sum_j S_{ij}^{pq}\left.\frac{\partial w}{\partial x^r}\right|_j, \quad (35)$$

where it is assumed that all quantities except the smoothing length are constant. Assuming a continuous expression for the smoothing length it is possible to rewrite Eq. 35 as an integral. From the momentum equation, Eq. 2, the acceleration should be zero since the pressure is constant. This is generally not true in an integral version of Eq. 35, neither in the case with varying shape of the kernel, nor with spherical kernels but variable smoothing length. The expression is not analysed, partly because the suggested scaling is not acceptable, partly because a few examples for certain values are not very enlightening. It suffices to note that spherical kernels give errors if the smoothing length varies over the kernel, and that a similar problem naturally occurs when a varying shape of the kernel is used. The errors do, however, become more severe, and at some degree of kernel deformation, it becomes physically meaningless to treat the fluid as a sum of two-particle interactions.

Another concern is the particle distribution, which according to the SPH-approximation is assumed to be isotropic. Using a spherical kernel on a cubic centered initial distribution, which for example Katz & Gunn (1991) did, there are directions in which particles will be found and others where there are no particles. Then a quantity does not satisfy the linear approximation, but additional terms are needed. There are to my knowledge no calculations on the validity of the linear approximation in a cubic centered grid or other particle distributions, and in a model when the particles move from their initial positions. However, the linear expansion can be expected to be a sufficient approximation for the cubic centered grid. From the figures in both Fulbright et. al. (1995) and Shapiro et. al. (1996) the particles are highly anisotropicly distributed, and piled upon each other in the direction of compression. Given the arguments above, it is most certainly not possible, neither using a spherical nor ellipsoidal kernel, to get an accurate approximation of a quantity in such a particle distribution.

# 6. Discussion

In SPH the acceleration in a point in space is defined as an integration over the kernel around the point in question. In the standard SPH approximation this volume is a sphere defined by the smoothing length. Here this relation is derived for an arbitrary shape of the kernel, as used by for example Martel et. al. (1995) in the discrete particle approximation. The relation includes the derivative of the smoothing length regarding the angles in spherical symmetry. To satisfy the momentum equation it is shown that the kernel has to be a symmetric even function around the point. This implies that the kernel can be described by at the most three parameters, such an ellipsoid, and that the tensor **H** Martel et. al. uses has to be diagonal.

The kernel is then defined to be symmetric in the $xy$-plane. Thus, it is decribed by its semiaxis in the $z$-direction, and a smoothing length that gives a spherical kernel of the same size as the ellipsoidal one. Calculations for some moderately deformed shapes of the kernel show that the extra term with the derivative of the smoothing length, Eq. (19), is large for a relatively moderate ellipsoidity, as Fig. 1 shows.

The errouneous extra term can, in principle, be corrected for. It does however affect the acceleration to change sign at rather moderate ellipsoidicities. The suggested straightforward corrected expressions therefore have singularities at these points. There may, however, be other possible solutions to correct for the deviation in the SPH approximation from the momentum equation.

In the integration different parts of the kernel contribute at different amounts to the total acceleration. This angular dependent contribution can be calculated for an ellipsoidal kernel and compared with its spherical counterpart. The problem with the comparison is that the uncorrected calculation is angular dependent in itself, and that the suggested correction is not acceptable. Nevertheless, this is sufficient in order to note that the noise in an ellipsoidal kernel due to imperfect particle distribution is higher relative to its spherical counterpart. To study this problem in more detail requires knowledge about the implications due to the particle distribution in itself, a topic not covered here.

SPH is often considered as a sum of two-particle interactions. This in contrary to an interation, or discretization over the neighbourhood of a particle. In a two-body interaction it is possible to average parameters such as the pressure or the smoothing length, to conserve properties such as energy, momentum or angular momentum. In an example all quantities are held constant except the smoothing length, which as other parameters is approximated with a linear expansion. Although not a simulation of the two body interaction situation, it exemplifies the problem with varying smoothing length. According to the momentum equation the calculation should give a zero acceleration, but is a function of the derivative of the smoothing length. This is a result equivalent with the earlier stated conclusion, that the kernel has to be an even function around the calculation point. The smoothing length is related to the third root of the density, and therefore, (provided the density dstribution is well behaved in the neighbourhood) the smoothing length should not vary much. In the case with varying ellipsoidities, the derivative of the smoothing length between the two particles in the interaction can be considerable.

Calculations of the two artificial viscous acceleration terms by Monaghan and Lattanzio (1985) give different dependencies on the smoothing length. This in turn gives different angular dependencies for the contributions to the viscous acceleration. Such a calculation is straighforward, although, due to the definition of the artificial viscosity, the odd terms can not be neglected in a similar way as in the derivation of Eq. 11. This means that the kernel is not an even function in the calculation of the artificial viscosity and introduces a non-linear term in the acceleration which could give an non-linear velocity distribution. In practice, however, this should probably not influence the calculation as a whole. Meglicki et. al. (1993) in their Appendix A calculate an estimation of the Reynold's number using the linearity properties of the velocity distribution. They do, however, not take into account that half of volume of the kernel by definition does not contribute to the artificial viscosity. Therefore they cancel the odd terms in the integration and erroneously exclude the linear term. Hence they calculate an overestimation of the Reynolds number based upon the quadratic velocity terms, which should not be considered in the linearization of the environment of a point in the fluid in the first place.

Without considering the problem of adapting the shape of the kernel to its environment, the consequences of a simple implementation of ellipsoidal kernels are discussed. It is shown that the method faces serious problems that have to be solved before any attempt to use it in any model. The theoretical aspects of SPH are not fully investigated, and some the difficulties discussed here probably originates from a vague description of the SPH method as a whole, a lack of test cases and a general preference of producing an efficient code rather than an efficient method.